\documentclass[twocolumn,showpacs,preprintnumbers,amsmath,amssymb]{revtex4-1}
\usepackage{graphicx}% Include figure files
\usepackage{dcolumn}% Align table columns on decimal point
\usepackage{bm}% bold math

\begin{document}

\preprint{APS/123-QED}

\title{Vacancy theory of melting}% Force line breaks with \\

\author{Leonid S. Metlov}

 \email{lsmet@fti.dn.ua}
\affiliation{Donetsk Institute of Physics and Engineering, Ukrainian
Academy of Sciences,
\\83114, R.Luxemburg str. 72, Donetsk, Ukraine
}%

\date{\today}% It is always \today, today,

\begin{abstract}
The features of alternative approach of non-equilibrium evolution thermodynamics are shown on the example of theory 
of vacancies by opposed to the classic prototype of Landau theory. On this foundation a strict theory of the melting 
of metals, based on development of Frenkel ideas about the vacancy mechanism of such phenomena, is considered. 
The phenomenon of melting is able to be described as a discontinuous phase transition, while the traditional Frenkel's 
solution in the region of low-concentration of vacancies can describe such transition only as continuous one. The problem 
of limiting transition of shear modulus to zero values in the liquid state, as well as the problem of the influence of 
extended state of vacancies on their mobility, is discussed. 
\end{abstract}

\maketitle

\section{Introduction}
The solids have a large number of variants of crystallographic structures and violations of these structures with such 
defects as vacancies, dislocations, grain boundaries etc. Under different severe external influences, such as radiation, 
severe plastic deformation and others they change the internal crystallographic and defect structures. To describe 
the changes the techniques of the theory of phase transitions is used, where the concept of order parameter or internal 
state variable is introduced.

The rudiments of the concept of internal state variable were offered yet Duhem in 1903 \cite{Duh03} (see also more detailed 
review in Ref. \cite{mm94}). In 1928 the idea was reanimated by Herzfeld and Rice for the account of internal structure 
evolution of polyatomic gas when they studied the dispersion and attenuation of sound \cite{hr28}. It got development in 
other numerous researches of that time \cite{r33,k33,lt36,ml37}. Supplemented with elements of rational mechanics, the 
concept took a completed and closed form in the works by Coleman and Gurtin \cite{cg67,cn92}. In the modern science, the 
line is developed too \cite{m99,m01,b04,tn04,ld05}. In the works by Landau the concept was developed anew, through introducing 
the internal state variable in the form of the order parameter, the theory of phase transitions was built by him at a 
high-quality level \cite{lk54,ll69}. The Landau approach has resulted in further development in the theories of the phase 
fields \cite{akv00,kkl01,esrc02,lpl02,rbkd04,gptwd05,akegny06,s06,rjm09}. In addition, the mesoscopic non-equilibrium 
thermodynamics approach was developing to fit the description of soft matter \cite{rrpm98,rg03,rvr05,pmrl11}. Studies of 
the problem gave birth to a plenty of variants of kinetic equations, describing the evolution of internal structure during 
non-equilibrium processes within the framework of different approaches (see \cite{m10} and review there).

Earlier, an alternative approach of non-equilibrium evolutional thermodynamics (NEET) proposed to describe the processes 
in solids with structural defects \cite{m10,m07,m07a,m08}. In Ref. \cite{m10} the problem was subdivided into thermal and structural parts. In the 
first part, the molecular dynamics simulation was used to represent a thermo-motion as a «superposition» of equilibrium and 
non-equilibrium contributions. The non-equilibrium constituent was the acoustic emission, arising from dynamic transition 
phenomena at a moment of defect (dislocations) formation. The acoustic emission has not been considered such context before. 
There it was considered not only as an attenuating wave process as a consequence of both scatter on internal heterogeneities 
and nonlinear pumping to other frequency regions of the spectrum, but as a part of thermodynamic process of the internal 
energy transformation. The resulting were kinetic (evolution) equations for the production of «non-equilibrium entropy» and 
its sinking into an equilibrium thermal subsystem.

The structural part was demonstrated on the example of a solid with vacancies, and the evolution equations were deduced in 
terms of the internal energy. In Ref. \cite{m11} this part was justified with the attraction of the elements of statistical physics, 
namely, the probability distribution function (PDF). There are two levels of rigor in the description of non-equilibrium 
processes. At the first level the PDF, which does not change in time is taken. Maxima of this function determine the 
equilibrium (most probability) states and the system tends to one of them starting from an arbitrary non-equilibrium one. 
The second level takes into account the possibility of probability evolution in time (kinetic equations of Boltzmann, 
Fokker-Plank etc). But already at first level of rigor one can obtain new additional relationships.

There are few approaches to description of the melting of solids. The most simple is based on the Lindemann criterion, 
which determines a melting-point by the amplitude of thermal vibrations. That is, when the amplitude becomes of the order 
of 10 percent of interatomic distance, the crystal lattice loses stability and collapses (melts) \cite{l1910,r69}. This 
approach is not, however, universal type and for different crystals of BCC-, FCC- and HCP-type, the Lindemann constant is 
different. In addition, the Lindemann theory is one-phase, it does not determine the free energy for the liquid state \cite{cb09}. 
The second approach is based on the mechanical Born criterion of stability, a solid melts when the shear modulus becomes 
zero \cite{b39,ml07}. However, the Born model is also one-phase theory which does not contain direct description of the liquid 
phase. The third approach, issuing from Frenkel and Eyring works, is based on order-disorder transitions with the participation of 
vacancies \cite{f55,er65,jkc75}. Unlike the former approaches this theory is biphase, however, up to now the discontinuous 
nature of transition between the solid and liquid state was not taken into account.

In this paper, within the framework of the developed NEET approach the vacancy theory of the melting of solids taking into 
account its discontinuous nature is considered. In part II, formulation of NEET is given for a solid with vacancies and 
interstitial atoms. For a homogeneous problem the PDF is obtained and a new form of kinetic equations symmetric in view of 
the use of free and internal energy is deduced. In the same part the vacancy theory of the melting of metals is considered 
from positions of Frenkel's concept. In part III, the discussion of the obtained results is given. The problem of the shear 
modulus tending to zero during transition to liquid media is discussed. Part V contains the summarizing conclusions.

\section{SOLID WITH VACANCIES AND INTERSTITIAL ATOMS}

\subsection{Probability distribution function}

The expressions for the thermodynamic probability $W_{V}$ and the configurational entropy $S_{cV}$ assuming the maximal 
degeneration of the microstates were obtained by Boltzmann for a solid with vacancies \cite{f55}
\begin{eqnarray}
\label{b1}
W_{V}=\dfrac{(N+n_{V})!}{N!n_{V}!},  \\
\label{b2}
S_{cV}=k_{B}\ln W_{V},
\end{eqnarray}
where $N$ is a total number of atoms in the crystal, $n_{V}$ is a number of vacancies, $k_{B}$ is the Boltzmann constant. 
Note that the configurational entropy $S_{cV}$ is a one-valued function of the number of vacancies and it does not depend on 
their energy (and on temperature as well).

The expression for the thermodynamic probability $W_{i}$ and the configurational entropy $S_{ci}$ for interstitial atoms can 
be quite similarly written down for the maximal degeneration in microstates
\begin{eqnarray}
\label{b3}
W_{i}=\dfrac{(N_{i}+n_{i})!}{N_{i}!n_{i}!},  \\
\label{b4}
S_{ci}=k_{B}\ln W_{i},
\end{eqnarray}
where $N_{i}$ is a total number of equilibrium positions for interstitial atoms in a crystal (local minima of the potential 
energy), $n_{i}$ is number of interstitial atoms. For different symmetries of crystals the number of such positions can be 
different and multiple to $N$. Note that the configurational entropy $S_{ci}$ is also a one-valued function of the number of 
interstitial atoms, which does not also depend on the energy of interstitial atoms and temperature. Provided that the 
microstates of vacancies and interstitial atoms are statistically independent, the total thermodynamic probability $W$ of 
the microstates is
  \begin{equation}\label{b5}
W=W_{V}W_{i}
  \end{equation}
and the total configurational entropy $S_{c}$ equals to the sum of entropies of subsystems
  \begin{equation}\label{b6}
S_{c}=S_{cV}+S_{ci}.
  \end{equation}

Total probability of the state containing $n_{V}$ vacancies and $n_{i}$ interstitial atoms will contain a limiting exponential 
multiplier, probability of Gibbs for this microscopic state \cite{s89,g97}
 \begin{equation}\label{b7}
f(n_{V},n_{i})=\dfrac{W}{Z}\exp(-\dfrac{U(n_{V},n_{i})}{k_{B}T}),
  \end{equation}
where $Z$ is the normalizing statistical sum, $U(n_{V},n_{i})$ is the internal energy taking into account the existence of 
vacancies and interstitial atoms, $T$ is the temperature. In conditions of the total degeneration the internal energy is 
a linear function of the defect number \cite{f55}
 \begin{equation}\label{b8}
U=U_{0}+u_{V0}n_{V}+u_{i0}n_{i},
  \end{equation}
where $U_{0}$ is the internal energy with no contribution of defects taken into account, $u_{V0}$, $u_{i0}$ are average 
energies of a vacancy and an interstitial atom, which in this case are constant values for all possible configurations of 
atoms of the system.

The condition of total degeneration is practically exactly satisfied for the low numbers (concentrations) of vacancies and 
interstitial atoms, when the interaction between defects can be neglected. At the same time, this condition is not fulfilled 
for those configurations for which vacancies are close to each other or merge at all (bi-vacancies, triple vacancies, vacancy 
pores) and at a high vacancy concentration. The same refers to the interstitial atoms as well. In these cases it is necessary 
to take into account the removal of condition of total degeneracy due to the interaction of point defects. As known, the 
coupling interaction reduces the total energy of a system that reduces the effect of action of the limiting exponential 
multiplier in Eq. \ref{b3} and results in a possible appearance of long-living configurations, which, thus, can compete with more 
numerous configurations of higher energy.

The probability distribution function with total (internal) energy $E_{l}$ is \cite{lsm11}
 \begin{equation}\label{b9}
f(E_{l})=\dfrac{w(E_{l})}{Z}\exp(-\dfrac{E_{l}}{k_{B}T}),
  \end{equation}
where $w(E_{l})$ is the distribution of states in energy $E_{l}$ or the number of microstates (configurations) with energy 
$E_{l}$, $Z$ is the statsum over all energy states of the system
 \begin{equation}\label{b10}
Z=\sum_{l=1}^{All states} w(E_{l}) \exp(-\dfrac{E_{l}}{k_{B}T}).
  \end{equation}

The states are numbered in the order of energy growth $E_{l+1} > E_{l}$. Distribution $w(E_{l})$ depends on the number of 
vacancies and interstitial atoms, and on their ratio, as well as on symmetry of their location (ordering). On this account, 
it is impossible to fix one-to-one correspondence between internal energy and number or concentration of defects. Finding of 
function $w(E_{l})$ is a hard combination problem, determination of power spectrum of $E_{l}$ as a set of acceptable energies 
for $l$ state, is also a stubborn problem. However, such one-to-one correspondence can be obtained for the average value of 
internal energy $U$ over all of the states for the fixed number of vacancies $n_{V}$ and interstitial atoms $n_{i}$. Using 
expression for PDF (\ref{b6}), it is possible to write down
 \begin{equation}\label{b11}
U(n_{V},n_{i})=\dfrac{1}{Z}\sum_{l=1}^{(n_{V},n_{i})} E_{l}w(E_{l}) \exp(-\dfrac{E_{l}}{k_{B}T}).
  \end{equation}

Here, unlike the case of Eq. (\ref{b8}), the internal energy is not the linear function of $n_{V}$ and $n_{i}$, but is 
the function of general type. The increase of fraction of the symmetric low-energy states diminishes the average value 
of the internal energy. The transition to the states, where high-concentration vacancies and interstitial atoms can pass 
to higher-energy extended (and mobiler) states \cite{gfs80,gfs83,tt07}, results in internal energy growth. To reflect all 
the properties of the internal energy it is presented in the form of a polynomial with alternating signs
\begin{eqnarray}\label{b12}
\nonumber
U=U_{0}+\sum_{m=V,i}^{} \sum_{k=1}^{\infty} \dfrac{(-1)^{k-1}}{k}u_{m,k-1}n_{m}^k- \\
-\sum_{k=1}^{\infty} \sum_{l=1}^{\infty} (-1)^{k+l}u_{kl}n_{V}^kn_{i}^l.
\end{eqnarray}

The first sum describes a contribution from every subsystem of point defects separately; the second one describes a 
contribution from the interaction between vacancies and interstitial atoms. Sign ``minus'' at the second sum is because 
the interaction between vacancies and interstitial atoms is, on the whole, of the attracting nature. Ignoring the degeneracy 
within one set of the states with the same number of vacancies $n_{V}$ and the number of interstitial atoms $n_{i}$, for such 
average internal energy we can write down the «effective» probability distribution function in the form of Eq. \ref{b7} with 
determinations (\ref{b1}), (\ref{b3}) and (\ref{b5}), in which the internal energy is set not by Eq. (\ref{b8}), but by a 
more general expression (\ref{b12}). Bringing the variables independent of the number of defects $n_{V}$ and $n_{i}$ into 
the ``inessential'' constant $1/Z$ we can write expression (\ref{b7}) in the form
 \begin{equation}\label{b13}
f(n_{V},n_{i})=\dfrac{1}{Z}\prod_{i=1}^{N}(n_{V}+i) \prod_{j=1}^{N_{i}}(n_{i}+j)\exp(-\dfrac{U}{k_{B}T}).
  \end{equation}

Note that the Boltzmann thermodynamic probability for vacancies and interstitial atoms is built within the concept of 
statistical independence of forming these subsystems (positions of interstitial atoms do not coincide with positions of 
lattice sites). At the same time, in the total distribution function these systems are not independent because of the mixed 
terms in the internal energy (see the last sum in (\ref{b12})).

Values of $n_{V}$ and $n_{i}$, at which PDF has extreme values (equilibrium states), can be found from the following 
transcendent equations \cite{m11}
\begin{eqnarray}\label{b14}
\sum_{i=1}^{N+n_{V}}\frac{1}{i}-\sum_{i=1}^{n_{V}}\frac{1}{i}-\frac{u_{V}}{kT_{B}}=0, \\
\label{b15}
\sum_{j=1}^{N+n_{i}}\frac{1}{j}-\sum_{j=1}^{n_{i}}\frac{1}{j}-\frac{u_{i}}{kT_{B}}=0.
\end{eqnarray}
where
\begin{eqnarray}
\label{b16}
\nonumber
u_{V}\equiv\frac{\partial U}{\partial n_{V}}=\sum_{k=1}^{\infty}(-1)^{k-1}u_{V,k-1}n_{V}^{k-1}- \\
-\sum_{k=1}^{\infty} \sum_{l=1}^{\infty} (-1)^{k+l}ku_{kl}n_{V}^{k-1}n_{i}^l \\
\label{b17}
\nonumber
u_{i}\equiv\frac{\partial U}{\partial n_{i}}=\sum_{k=1}^{\infty}(-1)^{k-1}u_{i,k-1}n_{i}^{k-1}- \\
-\sum_{k=1}^{\infty} \sum_{l=1}^{\infty} (-1)^{k+l}lu_{kl}n_{V}^kn_{i}^{l-1}
\end{eqnarray}
are energies or chemical potentials of a vacancy and an interstitial atom. Eqs. (\ref{b16}), (\ref{b17}) are the equations 
of state for a general non-equilibrium case. Eqs (\ref{b14}), (\ref{b15}) mesh due to the mixed terms in the expansion 
of the internal energy in (\ref{b12}), determining the interaction between the subsystems of vacancies and interstitial atoms.

The first two terms in Eqs (\ref{b14}), (\ref{b15}) are sums $S$ of slowly divergent harmonic series. This part of the 
equations depends only on system size and doesn't depend on material parameters. It is the fundamental part of Eqs. (\ref{b14}), 
(\ref{b15}), decreasing with the growing parameter $n_{V}$ or $n_{i}$, accordingly (curves 1 of Fig. 1, 2, 3). 
To calculate the fundamental curve $S$ we took $N=N_{i} = 2000$. The last terms in Eqs. (\ref{b14}), (\ref{b15}) depend 
on the parameters of material through the coefficients of $u_{k}$, and they are materiel parts (M) of Eqs. (\ref{b14}), 
(\ref{b15}). As seen, the positions of Eqs. (\ref{b14}), (\ref{b15}) roots coincide with the maxima of PDF, calculated directly 
by formulas (\ref{b12}), (\ref{b13}).

In the case of low concentration of vacancies and interstitial atoms Eqs (\ref{b14}), (\ref{b15}) can be solved analytically 
and it gives the well-known Frenkel solution \cite{f55}.

\subsection{Equilibrium states}

\begin{figure}
\hspace{0.06 cm}
\includegraphics [width=3 in] {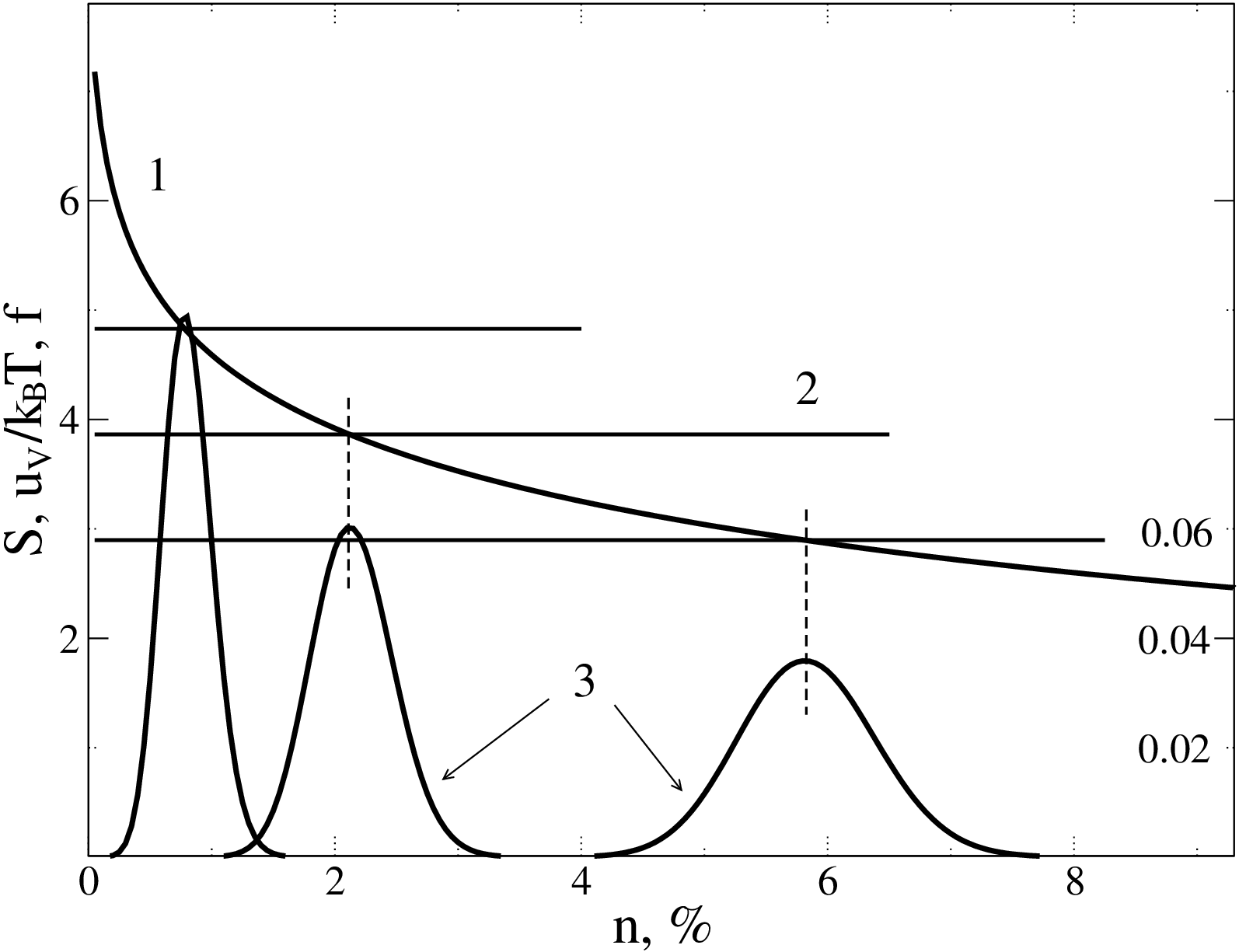}% Here is how to import EPS art
\caption{\label{f1} Solution of Eq. (\ref{b14}) for a constant energy of vacancy. The fundamental curve $S(1)$, 
the material curve (2) (proportional to energy of vacancy $u_{V}$) and PDF (3): $u_{V0}=1 (eV)$. Other parameters equal zero. 
$T=2400; 3000; 4000 (K)$. The right-hand scale is for PDF. Hereafter the graphs are constructed as a function of 
vacancies concentration in percent $n=100•n/N$.}
\end{figure}

It is known that the interstitial atoms have higher mobility as compared to vacancies, therefore they come to the equilibrium 
state first (adiabatic limit). For the times larger than the adiabatic limit the contribution of interstitial atoms can be 
ignored, as a result we get a pure vacancy problem \cite{m11}. Consider this problem in varying degree of approximation. In the 
linear with respect to the number of vacancies approach for the internal energy or for the constant vacancy energy there is 
only one solution of Eq. (\ref{b14}) that is the well-known Frenkel solution \cite{f55} (Fig. \ref{f1})
 \begin{equation}\label{b18}
n_{Ve}=N\exp(-\dfrac{u_{V0}}{k_{B}T}),
  \end{equation}
where $n_{Ve}$ is the equilibrium value of the number of vacancies.

At lowering the energy of vacancy or at the increase of temperature the solution is continuously displaced to region of 
higher number (or concentration) of vacancies. We can formally get any value of vacancy concentration, but, it is clear, 
that too high vacancy concentrations of about $100$ percent lose physical sense and the theory becomes useless. However 
Frenkel proposed to consider not too high vacancy concentrations of about $10$ persent as transition to the liquid state. 
At the same time, the solution proposed by him (in approximation of the independence of the vacancy energy of the vacancy 
concentration) can not explain or describe the process of melting of a solid as a jump-like first-order phase transition. 
Transition to the region of high concentration of vacancies is continuous during temperature growth.

\begin{figure}
\hspace{0.06 cm}
\includegraphics [width=3 in] {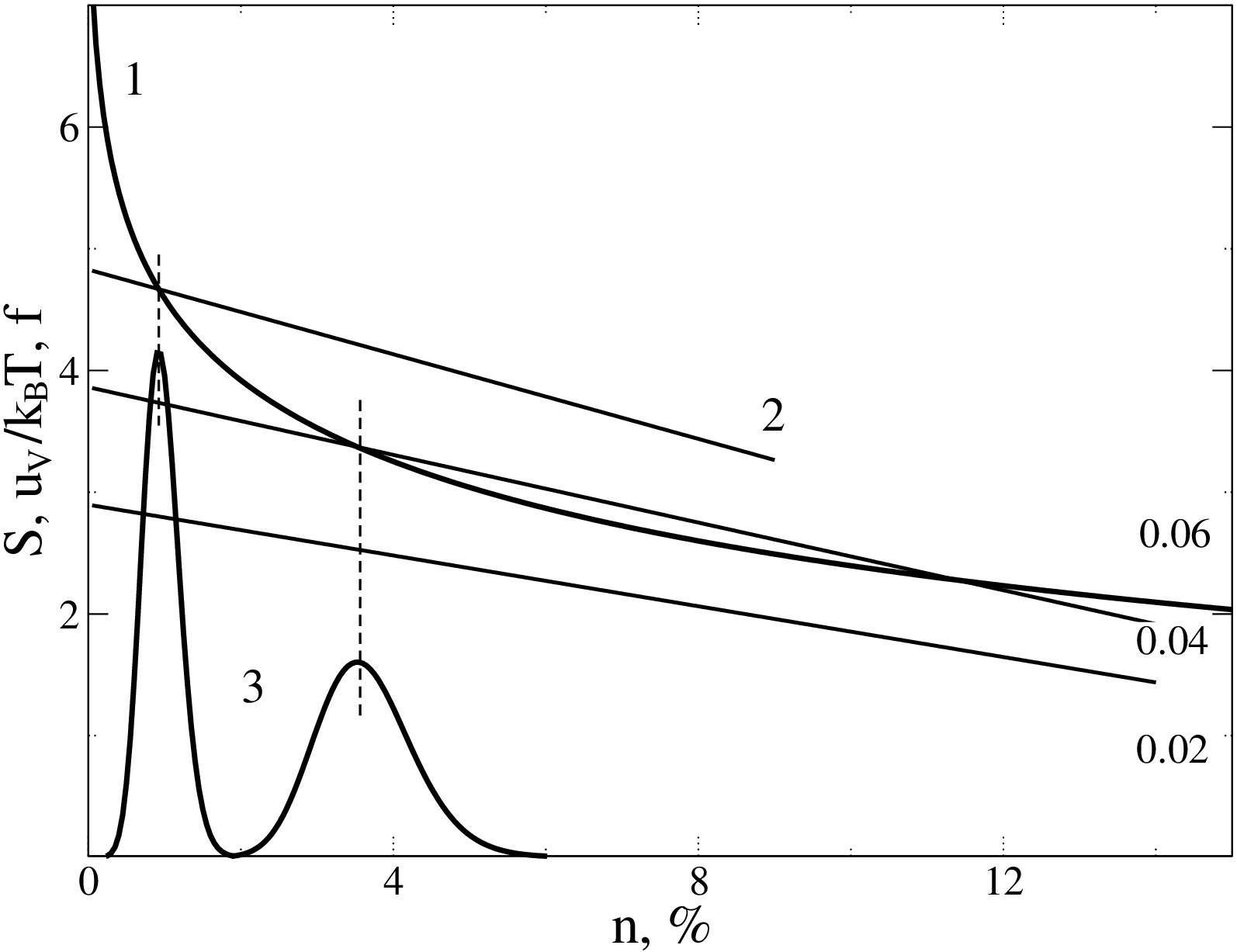}% Here is how to import EPS art
\caption{\label{f2} Solution of Eq. (\ref{b14}) for linear dependence of vacancy energy versus the vacancy concentration. 
The fundamental curve $S(1)$, the material curve (2) and PDF (3):  $u_{V0}=1 (eV)$, $u_{V1}=0.0018 (eV)$. 
Other parameters equal zero. $T=2400; 3000; 4000 (K)$. The right-hand scale for PDF.}
\end{figure}

In the quadratic with respect to the number of vacancies approximation for the internal energy or in the linear 
approximation for vacancy energy Eq. (\ref{b14}) has already two solutions in the region of high energy of vacancy or in the 
region of low temperatures (Fig. \ref{f2}). 
One of them in the region of a lower concentration of vacancies, actually, coincides 
with the Frenkel solution, but it is somewhat shifted due to the interaction of vacancies. The second solution in the 
region of a higher concentration of vacancies describes the equilibrium or stationary solution of the problem as well; 
however, it corresponds to a minimum of PDF and is unsteady. Probability of this state is lower not only as compared to 
the stable stationary state but also to any non-equilibrium state. In addition, for the low energy of vacancy or for high 
temperatures we enter in a region in which the solutions of Eq. (\ref{b14}) are absent at all. The last circumstance testifies 
unsuitability of the approximation in this region, so a higher approximation is needed.

\begin{figure}
\hspace{0.06 cm}
\includegraphics [width=3 in] {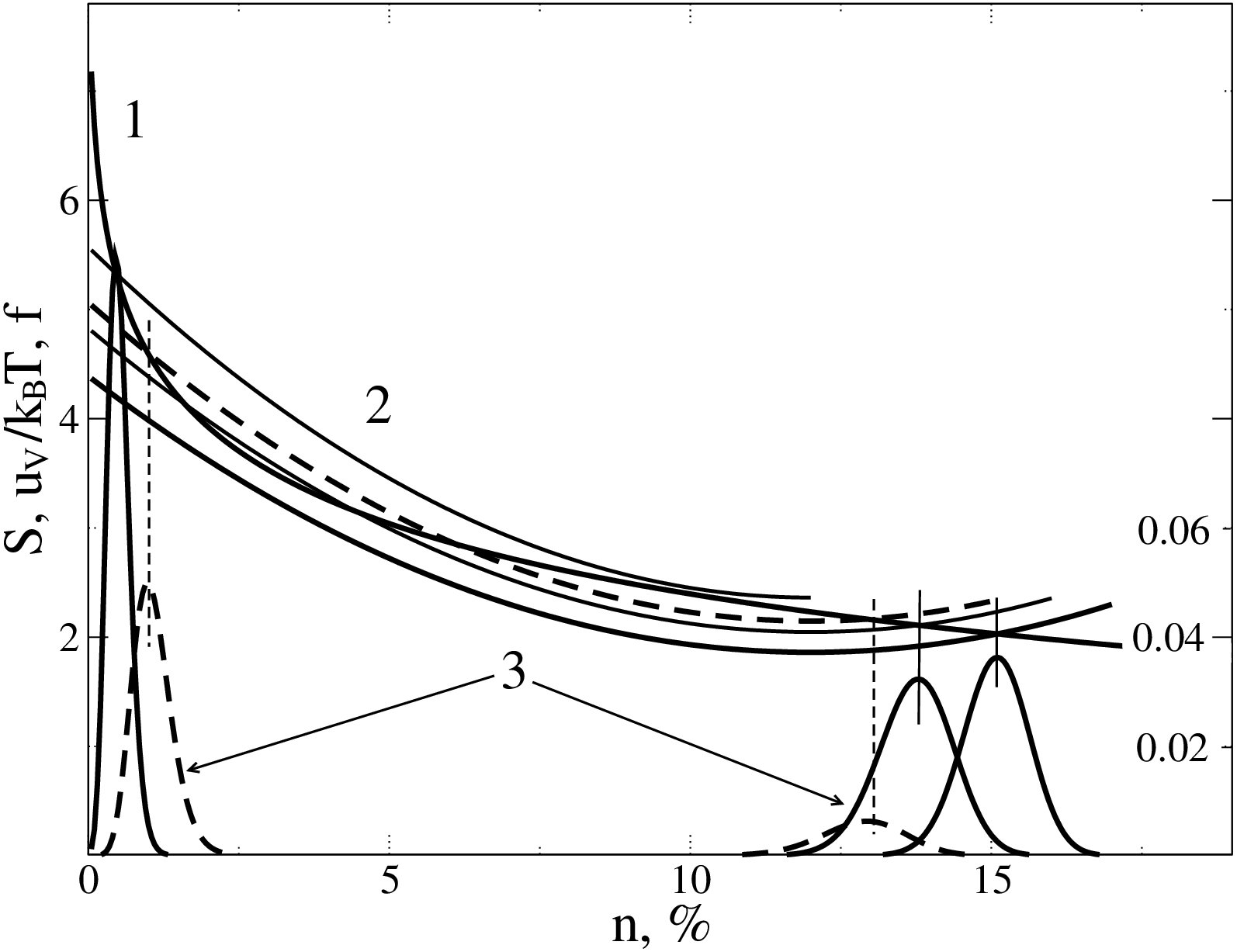}% Here is how to import EPS art
\caption{\label{f3} Solution of Eq. (\ref{b14}) for the quadratic dependence of vacancy energy on the vacancy concentration. 
The fundamental curve $S(1)$, the material curve (2) and PDF (3): $u_{V0}=0.125 (eV)$, $u_{V1}=0.6\cdot10^{-3} (eV)$, 
$u_{V2}=0.125\cdot10^{-5} (eV)$. $T=260; 286; 300; 330 (K)$. The case of $T=286 K$ is shown by dotted line. 
The right-hand scale is for PDF.}
\end{figure}

In the cubic with respect to the number of vacancies approximation for the internal energy or in the quadratic 
approximation for the energy of vacancy, Eq. (\ref{b14}) can already has one or three solutions (Fig. \ref{f3}). 
In region of high values of the energy of vacancy or low temperatures Eq. (\ref{b14}) has one solution, which is a Frenkel 
solution modified due to the nonlinear contributions. At lowering the energy of vacancy or with temperature growth the 
equation has three solutions, one of which (the left-hand) is the modified Frenkel solution, the second one (intermediate) 
is unsteady, and the third one (the right-hand) in the region of high concentration of vacancies can be treated as one 
proper to the liquid state of the matter. In this case, we have an equilibrium coexistence of solid and liquid phases of 
the matter. For a still greater increase of temperature we pass to the region with only one solution, which corresponds 
to clear-melted matter. The transition from the solid state to liquid one is a jump-like first-order phase transition. 
In order to talk about the transition just to the liquid state (not for example to the amorphous state), we need not 
only the high concentration of vacancies, but also the shear modulus of material going to zero. This issue will be 
discussed in the next part in detail.

Influence of the interstitial atoms can be qualitatively estimated understanding that they have higher energy as 
compared to vacancies, and therefore in the equilibrium state have more low concentration. As, with (\ref{b13}) and (\ref{b17}), 
the sign ``minus'' is at the constant $u_{11}$, then, due to cross effects, this results in diminishing of the energy of 
vacancy $u_{V}$, and to some displacement of the roots of the equation (\ref{b14}) to the region of a higher number 
(concentration) of vacancies. For the case of interstitial atoms to be directly participating in the processes of 
melting the equation (\ref{b15}) should have one more solution. As their energy is higher, their material curve will 
be considerably higher than that for vacancies. Consequently, while the equation (\ref{b14}) has two solutions, the 
equation (\ref{b15}) has only one solution, and the interstitial atoms will not directly and independently participate in 
the processes of melting, but will only modify the participation of vacancies a little.

\subsection{Set of non-equilibrium thermodynamic potentials}

The free energy $F=U-TS$, where $S$ is the entropy, was introduced for solution of quasi-equilibrium thermal problems 
in the defect-free condensed matter, as one of types of thermodynamics potential, along with the internal energy, 
enthalpy and Gibbs energy. In certain external conditions it, as well as other thermodynamic potentials for other 
conditions, possessed extreme properties as opposed to internal non-equilibrium processes. For imperfect solids, 
in particular solids with vacancies, according to the hypothesis of Boltzmann, the free (configurational) energy was 
introduced by analogy
 \begin{equation}\label{b19}
F_{cV}=U-TS_{cV},
  \end{equation}
where configurational entropy $S_{cV}$ is set by Eqs (\ref{b1}), (\ref{b2}). Properties of thermodynamics potential were also 
attributed to this (configurational) free energy, and it possesses the property of minimality as to a variation of 
concentration of structural defects. It is easy to see that this property issues from the fact that the free energy, in a 
logarithmic scale and with an opposite sign, coincides within a constant with PDF (\ref{b7})
 \begin{equation}\label{b20}
F_{cV}=-k_{B}T \ln{(Zf)},
  \end{equation}
and as a consequence the minima of the free energy automatically coincide with the maxima of PDF (compare curves 1 
and 2 of Fig. 4), which determine the most probable, that is, equilibrium states. In other words, the free energy 
$F_{cV}$ is a quantity reciprocal to PDF, expressed in energy units on a logarithmic scale.

Energies $F$ and $F_{cV}$ to some extent similar have important principal distinctions. Namely, it is known that 
the internal and free energies, related by the Legendre transformations, depend on the arguments of its own 
$U=U(T,V)$, $F=F(S,V)$, we name them symbolically the eigen-arguments of these potentials. This property and the 
respective differential relationships, allow to interpret these energies as thermodynamics potentials. If we compare 
the energy of vacancy $u_{V}$ to the temperature $T$, then in this context, it is necessary to confront the number 
of vacancies $n_{V}$ with entropy $S$. It would seem the configurational free energy $F_{cV}$ must be a function of vacancy 
energy $u_{V}$, however, from relationships (\ref{b1}), (\ref{b2}) and (\ref{b19}) it is follows that it is a function 
of the number or concentration of vacancies $n_{V}$. It can be concluded that the free energy $F_{cV}$ is not a 
thermodynamics potential in strict sense. 

\begin{figure}
\includegraphics [width=2.9 in]{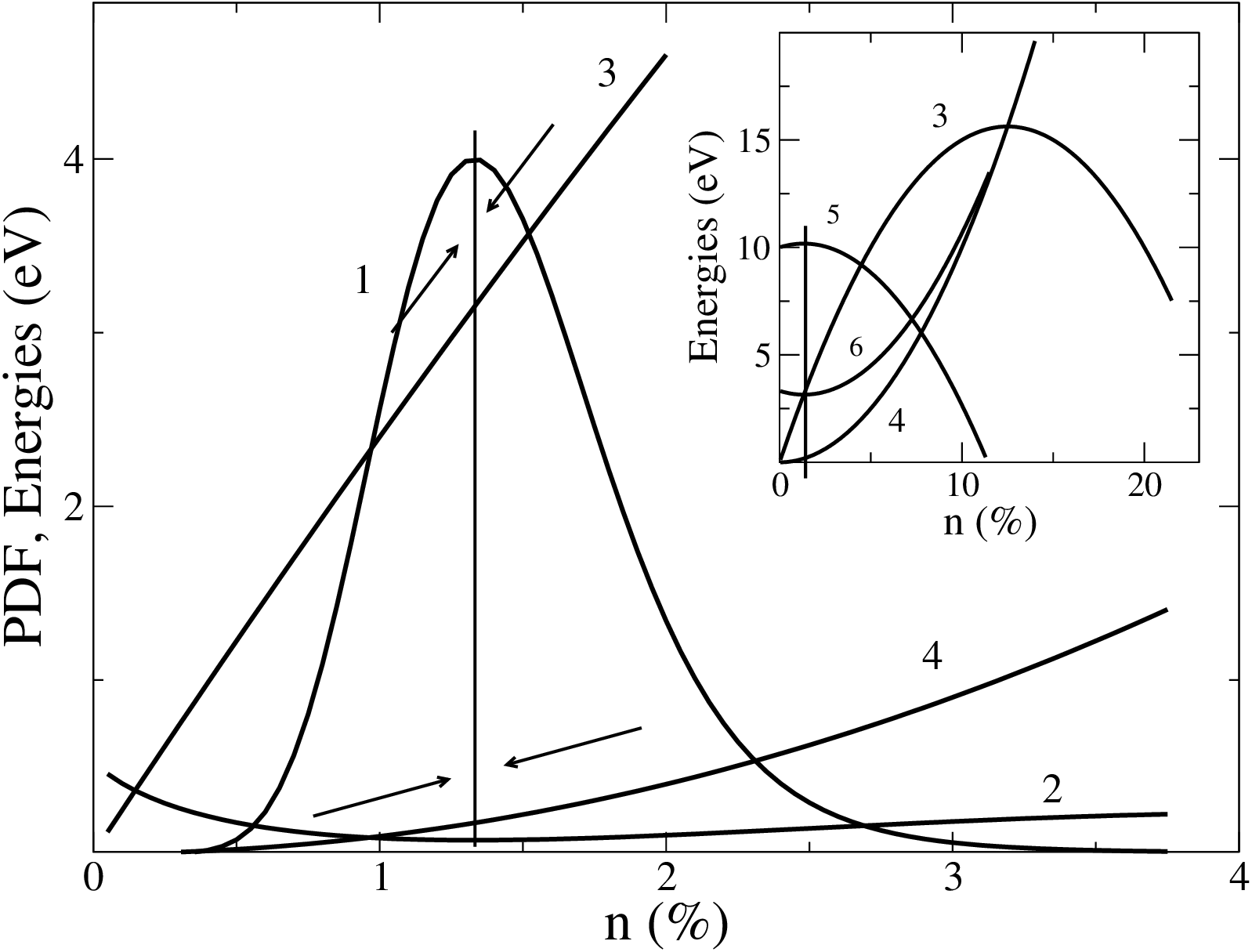}% Here is how to import EPS art
\caption{\label{f4} Basic set of thermodynamic functions for quadratic approximation: 1 -- probability distribution 
function $f$; 2 -- classic configurational free energy $F_{c}$; 3 -- internal energy $U$; 4 -- a tangent to $U$ 
at the equilibrium point; 5 -- modified configurational free energy $\tilde{F_{c}}$; 6 -- effective internal 
energy $\bar{U}$; 7 -- effective configurational free energy $\bar{F_{c}}$. The right-hand scale is for PDF. 
Straight line corresponds to equilibrium state.}
\end{figure}

To construct some thermodynamics potential of the free energy, note that in the case of vacancies according to (1) and (2), 
the configurational entropy $S_{cV}$ and the number of vacancies $n_{V}$ are in a mutual one-valued dependence with 
each other, that is, the number of vacancies $n_{V}$ can always be used as an independent argument instead of the 
configurational entropy $S_{cV}$. In this case, executing transformation of Legendre type with the pair of the 
conjugated thermodynamic variables $u_{V}$ and $n_{V}$, we get the alternative type of the free energy 
(curve 4 of Fig. \ref{f4})
 \begin{equation}\label{b21}
\tilde{F}_{cV}=U-u_{V}n_{V}.
  \end{equation}

It is easy to show 
 \begin{equation}\label{b22}
n_{V}=-\dfrac{\partial\tilde{F}_{cV}}{\partial u_{V}}.
  \end{equation}

Relationships (\ref{b16}) and (\ref{b22}) are the joint pair of equations for the internal energy $U$ and the 
modified configuration free energy $\tilde{F}_{cV}$ on one side, and for the vacancy concentration $n_{V}$ and vacancy 
energy $u_{V}$ on the other side. It can be seen that the vacancy concentration is an eigen-argument of the internal 
energy, and the energy of defects is an eigen-argument of the modified configurational free energy. In the quadratic 
approximation this dependence can be written in an explicit form (see curve 4 of fig. \ref{f4})
 \begin{equation}\label{b23}
\tilde{F}_{cV}=U_{0}+\dfrac{1}{2u_{V1}}(u_{V0}-u_{V})^2.
  \end{equation}

As the energy necessary for the formation of a new vacancy is lower, in the presence of other defects, than in a 
vacancy-free crystal, the quadratic correction in (\ref{b12}) has negative sign. Note that expression (\ref{b12}) 
is suitable for both the equilibrium and non-equilibrium state. In this approximation the internal energy is a convex 
function of the vacancy number, having a maximum at point $n_{V}=n_{Vmax}$, as is shown in the insert of Fig. \ref{f4}. 
In that approximation the modified configurational free energy is a concave function with a minimum at point 
$n_{V}=0$ (or $u_{V}=u_{0}$).

With relationships (\ref{b16}) and (\ref{b22}) it is easy to show that the stationary state corresponds to neither 
the maximum of the internal energy $U$ nor to the minimum of the free energy $\tilde{F}_{cV}$, as the stationary 
state is at point $n_{V}=n_{Ve}$, where $u_{Ve}$ and $n_{Ve}$ satisfy equations
  \begin{equation}\label{b24}
u_{Ve}=\dfrac{\partial U}{\partial n_{Ve}}\neq0,  
\quad n_{Ve}=-\dfrac{\partial \tilde{F}_{cV}}{\partial u_{Ve}}\neq0.
  \end{equation}

\begin{figure}
\hspace{0.06 cm}
\includegraphics [width=3 in] {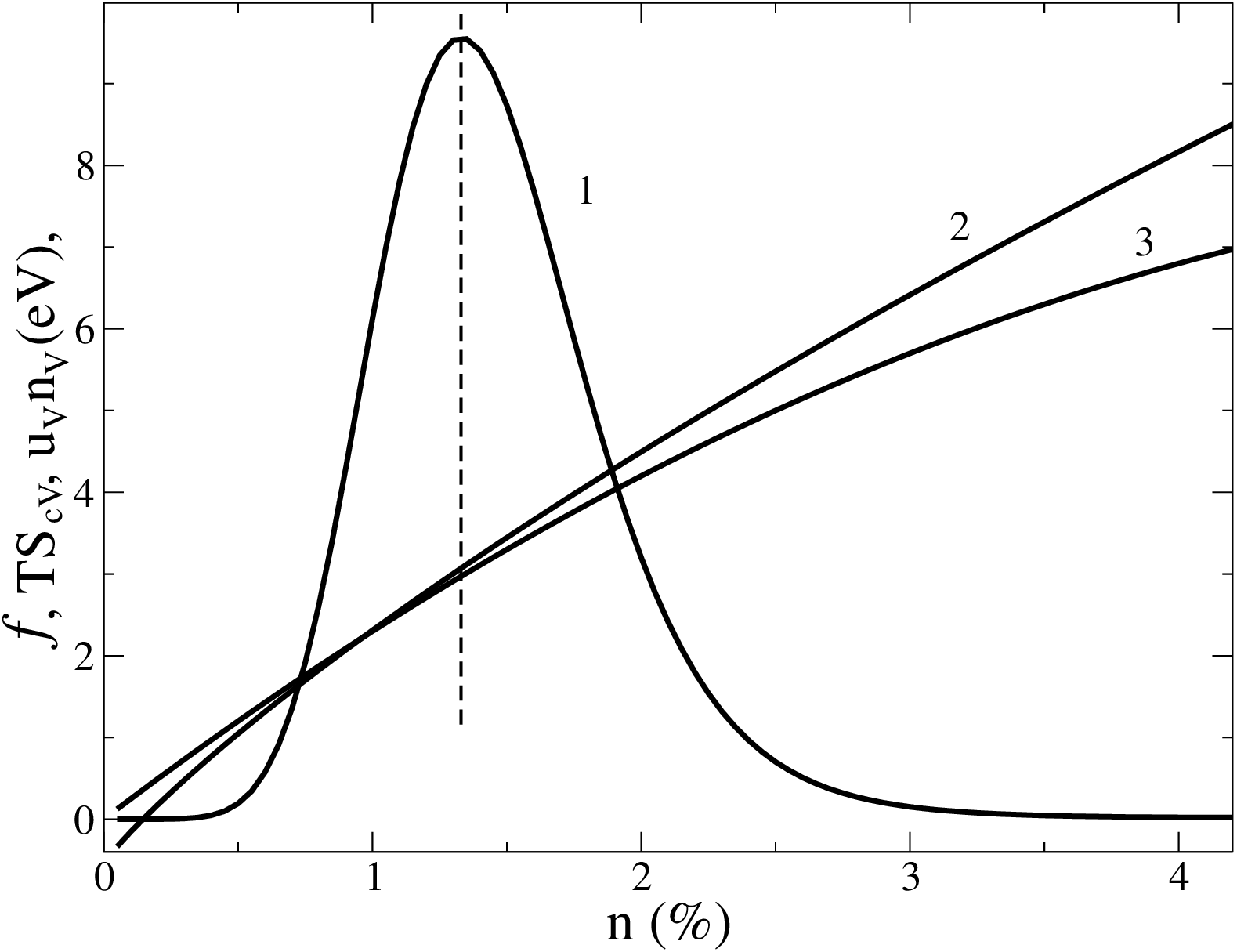}% Here is how to import EPS art
\caption{\label{f5} ``Bound energy'' of a solid with vacancies: curve 1 – PDF; curve 2 – bound energy in form $TS_{cV}$; 
curve 3 – bound energy in form $u_{V}n_{V}$.}
\end{figure}

It is known from thermodynamics that the product $TS_{cV}$ in (17), entered in determination of the canonical free 
energy $F_{cV}$, makes sense of the bound energy lost for the production of work. On the other side, the total energy 
of defects $u_{V}n_{V}$, in the main part, physically is the energy which is also lost for work production. Only a small 
part of it remains for work production. Then we can conclude that these two energies must be equal, at least, in the 
region of the equilibrium state (Fig. \ref{f5}).
 \begin{equation}\label{b25}
TS_{cV} \approx u_{V}n_{V}.
  \end{equation}

The energies are equivalent as we have deal, actually, with energy of the same origin, but written in different forms. 
The quantities, included in this relationship, are exactly enough defined, and energy $u_{V}n_{V}$, is to be understood 
as the total energy stocked in the defects. Temperature $T$ is the temperature of thermomotion, and it is directly unconnected 
with a defect subsystem. Approximate equality (23) reflects a dynamic thermal equilibrium between the thermal subsystem, 
as a dynamic motion of atoms, and the static subsystem of defects. The configurational entropy $S_{cV}$ is an interface 
between these subsystems. It is also interesting to note that both temperature $T$ and energy of vacancy $u_{V}$ are 
specific energetic characteristics of each subsystem. In addition, both the configurational entropy $S_{cV}$, and defect 
concentration $n_{V}$ possess property of additiveness, that along with their mutual one-valued accordance (\ref{b1}), 
(\ref{b2}) can be foundation for their interchangeability in general thermodynamic relationships.

Summarizing the above said, we can apply similar relationships and reasoning to interstitial atoms. The question of their 
application to other types of defects (dislocations, grain boundaries etc) remains open, as there are no simple and reliable 
statistical expressions for them, analogous to those presented in part II. At the same time, it is obvious that from the 
thermodynamics point of view all structural defects are characterized in identical way by surplus energy, spent for their 
formation, and they can be described by similar evolution equations, but written, for the present, at phenomenological level.

\subsection{Kinetic (evolution) equations}

If the system has deviated from the equilibrium (stationary) state it will tend to return back to that state at a speed 
the higher the larger the deviation. Evolutional equations for a solid with point defects, in this case, are 
\cite{m10,m11}
\begin{eqnarray}
\label{b26}
\nonumber
\dfrac{\partial n_{m}}{\partial t}=\pm\gamma_{n_{m}}(\dfrac{\partial U}{\partial n_{m}}-u_{me}),  \\
\quad \dfrac{\partial u_{m}}{\partial t}=\mp\gamma_{u_{m}}(\dfrac{\partial \tilde{F}_{cm}}{\partial u_{m}}+n_{me}).
\end{eqnarray}
Hereafter $m=V,i$

The form of kinetic equations (\ref{b26}) is symmetric with respect to the use of internal and configurational free energy. 
If the equilibrium state is closer to the maximum of internal energy, then, in view of solution stability, the upper sign 
is chosen (convex function). If it is near the minimum of internal energy, the lower sign is taken (concave function). 
Both variants of the kinetic equations are equivalent and their application is a matter of convenience.

At the same time, for a number of reasons, more preferable are evolution equations expressed in terms of the internal energy. 
First, concentration (or number) of defects is a more easily measurable variable, second, the internal energy is a base 
quantity in thermodynamics (the first law of thermodynamics), it is universal both for the equilibrium and non-equilibrium 
state and, finally, internal energy is the analogue of Hamiltonian. Setting the internal energy, as a function of state 
variables we specify the system and fully determine it from the thermodynamics point of view.

If we assume that the equilibrium energy of defect $u_{V}=u_{Ve}$ (or $u_{i}=u_{ie}$) and the number of defects 
$n_{V}=n_{Ve}$ (or $n_{i}=n_{ie}$) are slowly changing during an external load, then their product can be introduced 
under differentiation sign in (\ref{b26}). It permits us to specify a new kind (shifted) of internal and free (effective) 
energies (see curves 5 and 6 in Fig. \ref{f4}):
  \begin{equation}\label{b27}
\bar{U} =U-u_{me}n_{m},  
\quad \bar{F}_{cm} =\tilde{F}_{cm}+u_{m}n_{me}.
  \end{equation}

Then equations (\ref{b26}) are simplified a little
  \begin{equation}\label{b28}
\dfrac{\partial n_{m}}{\partial t}=\pm\gamma_{n_{m}}\dfrac{\partial \bar{U}}{\partial n_{m}},  
\quad \dfrac{\partial u_{m}}{\partial t}=\mp\gamma_{u_{m}}\dfrac{\partial \bar{F}_{cm}}{\partial u_{m}}.
  \end{equation}

The original potentials $U$ and $\tilde{F}_{cm}$ are connected by means of transformation of the Legendre type  (\ref{b21}). 
The shifted potentials $\bar{U}$ and $\bar{F}_{cm}$ are connected by means of transformation
  \begin{equation}\label{b29}
\bar{F}_{cm}=\bar{U}-u_{m}n_{m}+u_{me}n+u_{m}n_{me},
  \end{equation}
which differs from the Legendre-type transformation by anticommutation bracket $[u_{m}n_{m}]=u_{me}n_{m}+u_{m}n_{me}$.

For the shifted potentials the stationary point coincides with maximum of $\bar{U}$ (curve 5 of Fig. \ref{f4}) and with 
minimum of $\bar{F}_{cm}$ (curve 6 in Fig. \ref{f4}) for the upper signs in Eq. (\ref{b28}). Thus, $\bar{U}$ is an effective 
thermodynamic potential, for which the tendency of the original part of the internal energy to a minimum is completely 
compensated by the entropic factor. Free energy $\bar{F}_{cm}$ tends to a minimum, but it tends to a minimum in the space 
of eigen-argument $u_{m}$, while the free energy $F_{c}$ tends to a minimum in the space of ``non-eigen-argument'' $n_{m}$.

\begin{figure}
\includegraphics [width=2.9 in]{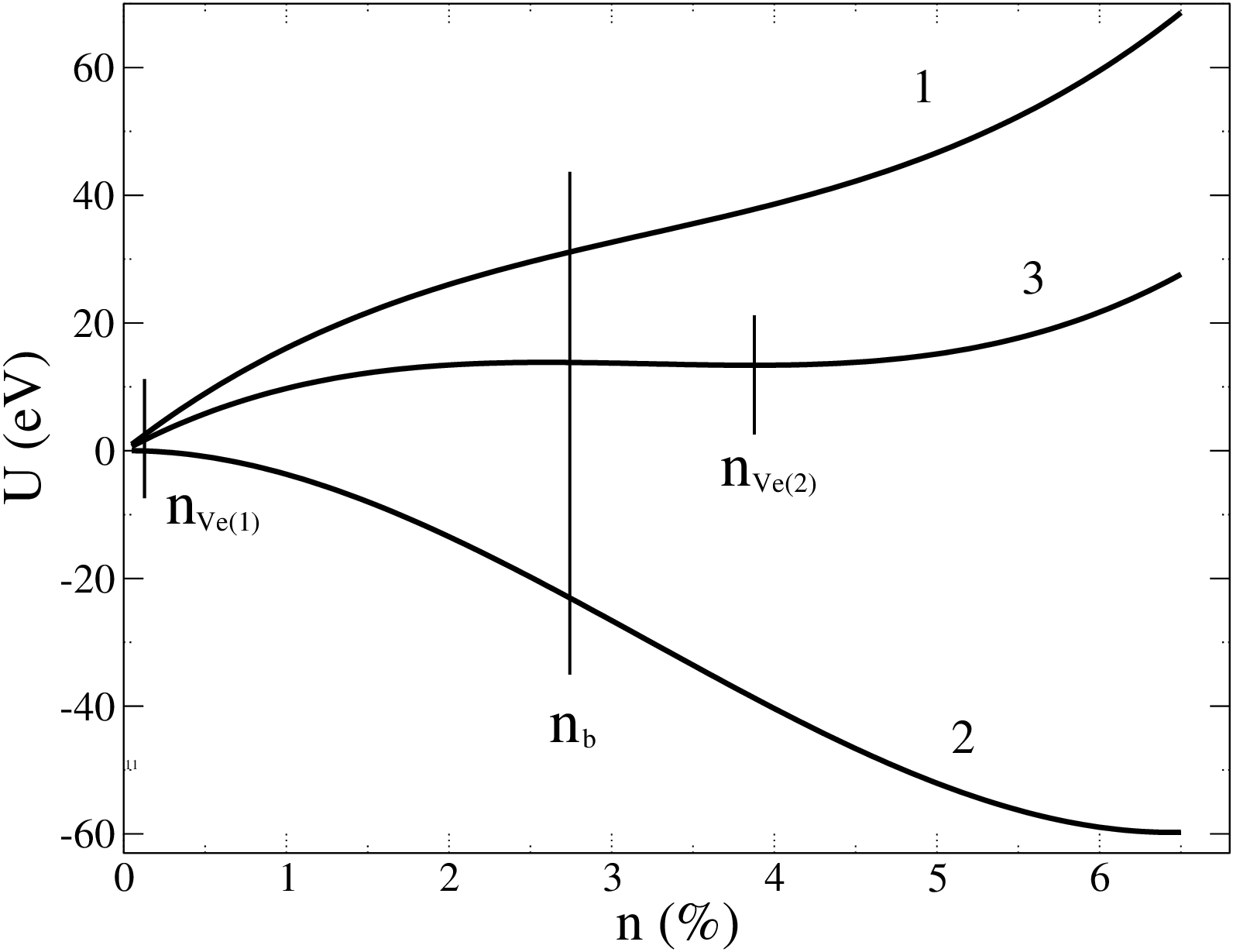}% Here is how to import EPS art
\caption{\label{f6} Internal (1) and effective internal energy (2, 3). $n_{Ve(1)}$ and $n_{Ve(2)}$ are vacancy 
concentration for the first and second stable equilibrium state, $n_{b}$ is an inflection point of the internal 
energy (1). For calculations, parameters were chosen close to parameters of copper $u_{V0}=1 (eV)$, $u_{V1}=0.022 (eV)$, 
$u_{V2}=0.00017 (eV)$.}
\end{figure}

In the presence of only one equilibrium state $n_{me}$, $u_{me}$, the effective internal energy is determined simply. 
In the presence of two such states transformation (\ref{b27}) can done for each one. In this case, we get two graphs of 
the effective internal energy (Fig. \ref{f6}).

As the first state is in the region of convexity of the internal energy (curve 1), the graph of its respective effective 
internal energy has a maximum (curve 2), and in equations (\ref{b26}), (\ref{b28}) the upper signs are chosen. 
The second state is in region of concavity of the internal energy, and the graph corresponding to its effective internal 
energy has a minimum (curve 3), and in equations (\ref{b26}), (\ref{b28}) lower signs are chosen. Obviously, the 
evolution of the system from non-equilibrium states to the nearest equilibrium states till the inflection point on 
the graph of internal energy should be described by evolution equations (\ref{b28}), taken with upper signs, and past 
at that point, vice versa, with lower ones.

\section{DISCUSSION AND PROSPECTS}

Above it has been shown, that with the increase in the temperature of a solid some new state with high concentration 
of vacancies is formed, which in accordance with the hypothesis of Ya. I. Frenkel can be interpreted as the liquid state. 
At the same time, according to the Born hypothesis a distinctive property of liquid is the absence of shear rigidity or 
equality to the zero of the shear modulus of material. Shear rigidity displays activation character, and it is in inverse 
dependence with mobility of point defects.

To study the issue let us consider a solid having only vacancies, by setting the internal energy in the form
  \begin{equation}\label{b30}
u(n_{V})=u_{0}+u_{V0}n_{V}-\dfrac{1}{2}u_{V1}n_{V}^2+\dfrac{1}{3}u_{V2}n_{V}^3,
  \end{equation}
where $u_{0}$ , $u_{Vl}$ ($l=1, 2, 3$) are some coefficients dependent on temperature $T$ and elastic deformation 
$\epsilon_{ij}^e$, as on control parameters. For the sake of convenience and further standardization we have passed 
from the number of vacancies to their concentrations $n_{V}$ ($n_{V}=n_{V}/N$), and in a similar way to specific 
internal energy $u=U/N$. Then
\begin{eqnarray}
\label{b31}
\nonumber
u_{0}=\dfrac{1}{2}\lambda(\epsilon_{ii}^e)^2+\mu(\epsilon_{ij}^e)^2,  \\
\quad u_{V0}=u_{V0}^*+\dfrac{1}{2}\bar{\lambda}(\epsilon_{ii}^e)^2+\bar{\mu}(\epsilon_{ij}^e)^2,
\end{eqnarray}
where $\lambda$ and $\mu$ are Lame moduli for a vacancy-free solid, $\bar{\lambda}$, $\bar{\mu}$ are corrections to the 
elastic moduli due to existence of vacancies (defect of a modulus), $\epsilon_{ij}^e$ is the first invariant of the 
tensor of deformations, $(\epsilon_{ij}^e)^2=\epsilon_{ij}^e\epsilon_{ji}^e$. It is assumed that contribution of 
invariants is the most substantial at the lower degrees of presentation (\ref{b30}).

Expression for the effective shear modulus within the framework of this model is
  \begin{equation}\label{b32}
\mu_{ef}=\mu-\bar{\mu}_{V}n_{V}.
  \end{equation}

The sign ``minus'' is chosen provided, that the effective modulus in material with vacancies is less than in the 
vacancy-free one (see Fig. 2.3 in \cite{mou04}). The first term $\mu$ is shear modulus in a vacancy-free crystal. 
It can be interpreted as a local (microscopic) modulus of the lattice in regions far from vacancies. The effective 
(macroscopic) modulus $\mu_{ef}$ is a quantity averaged over volume containing a lot of vacancies. It is remarkable, 
that at a phase transit point at melting, namely, the effective macroscopic modulus goes to zero exactly, while the 
microscopic modulus does not convert to zero \cite{ml07}.

A few types of the temperature dependence for the shear modulus are known. For example, there is a quadratic 
dependence $\mu(T)=\mu{0}(1-T/T_{0})$ \cite{ml07}. Other, more known semi-empirical relationship is \cite{v70,cg96,b05}
 \begin{equation}\label{b33}
\mu(T)=\mu_{0}-\dfrac{D}{\exp{\dfrac{T_{0}}{T}}-1},
  \end{equation}
where $\mu_{0}$ is the shear modulus for $T=0 K$, and $D$, $T_{0}$ are material constants. The constant $T_{0}$ can 
be treated as an activation energy $T_{0}=E_{a}/k_{B}$, and the relationship (\ref{b33}) is presented as
\begin{equation}\label{b34}
\mu(T)=\dfrac{\mu_{0}-(\mu_{0}+D)\exp{(-\dfrac{E_{a}}{k_{B}T}})}{1-\exp{(\dfrac{E_{a}}{k_{B}T}})}.
  \end{equation}

According to Born principle, the temperature of melting can be defined on condition $\mu(T_{m})=0$
\begin{equation}\label{b35}
T_{m}=\dfrac{E_{a}} {k_{B}\ln{\dfrac{\mu_{0}+D}{\mu_{0}}}}.
  \end{equation}

A formal agreement between the theory presented in part II and relationship (\ref{b35}) can be obtained by 
taking the temperature of phase transition for the temperature of melting $T_{m}$. At the same time, it is more 
logically to present the shear modulus as a two part one consisting of two terms, actually, shear modulus at zero 
temperature $\mu_{0}$ and an activation depressed term
\begin{equation}\label{b36}
\mu(T)=\mu_{0}(1-\exp{(-\dfrac{E_{a}}{k_{B}T}})).
  \end{equation}

However, in obedience to this expression, the shear modulus does not go to zero at limit temperature $T_{m}$, 
but only at $T=\infty$. It is known in the theory of amorphous materials, that activation energy is proportional to 
the shear modulus $E_{a}\sim\mu$ \cite{r11}. It can be assumed that a similar supposition is just for crystalline solids 
too, at least, in the pre-melting state, for in this state its structure becomes amorphous-like. Then relationship 
(\ref{b36}) can be rewritten as
\begin{equation}\label{b37}
\mu_{ef}=\mu_{0}(1-\exp{(-\dfrac{a\mu_{ef}}{k_{B}T}})).
  \end{equation}

It is taken into account that here we have the effective shear modulus.

It is remarkable that this equation has a trivial zero solution $\mu_{ef}=0$. This solution is important as it can 
serve for description of non-activation motion of liquid phase to which all solids pass as a result of melting.

\begin{figure}
\includegraphics [width=2.9 in]{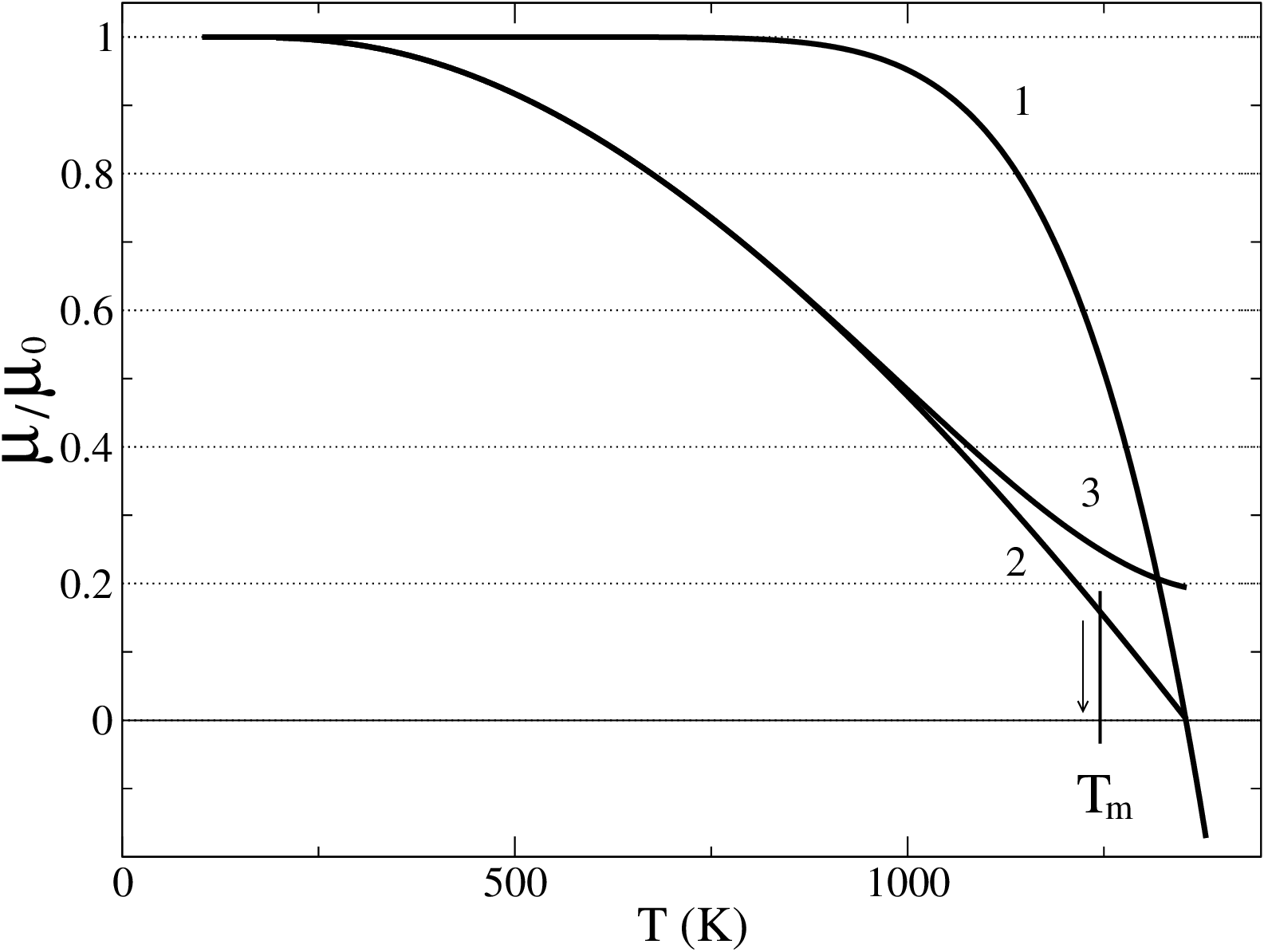}% Here is how to import EPS art
\caption{\label{f7} Temperature dependence of the shear modulus: curve 1 – semi-empirical shear modulus with 
(\ref{b34}); curve 2 – effective modulus with (\ref{b37}); curve 3 – shear micro-modulus. Arrow indicates the jump of 
the shear modulus to zero value at the real melting point.}
\end{figure}

The graphs of temperature dependences of the shear modulus are shown in fig. \ref{f7}. 
Curve 1 proper to semi-empirical theory (\ref{b34}), slowly changes at low temperatures and sharply diminishes to zero, 
passing to the region of negative unphysical values. Curve 2, built for the effective shear modulus by formula (\ref{b37}), 
changes already at low temperatures and is more declivous. It smoothly tends to zero, not passing to the region of 
negative values. In addition, zero values of the effective shear modulus identically satisfy equation (\ref{b37}), 
describing a liquid phase legalistically in the whole of temperature interval, however, this solution is realized 
only after the phase transition. Curve 3, which corresponds to the shear micro-modulus, does not go to zero at a 
``melting'' point.

The graphs of Fig. \ref{f7} correspond a continuous change with the temperature of the shear modulus in the solid 
phase. However, at the real melting-point because of the first-order phase transition the change can be sharper 
(see the arrow). Similar transitions are observed in the real materials (see, for example, Fig. 1 in Ref. \cite{fg05}).

At low temperatures, a basic contribution to shear deformation is from dislocations in view of their greater spatial 
size and higher mobility. In obedience to the Cottrell theory, mobility of dislocations is the higher, the wider their 
dislocation cores \cite{k53}. Mobility of vacancies at low temperatures is negligibly small.

At more high temperatures the situation, however, changes cardinally. Vacancies due to the thermomotion become 
delocalized in space, forming the so-called extended defects of higher mobility \cite{gfs80,gfs83,tt07}. The total 
energy barrier of a vacancy splits on a great number of barriers of lower height, which are considerably easy overcome 
due to thermal fluctuations. The number of variants of overcoming these barriers increases thus, here the vacancies have 
certain advantage over dislocations. The extended vacancy is a three-dimensional defect and the number of variants 
for ``lateral'' passing by potential barriers grows quadratically, while dislocation is a two-dimensional object and the 
number of variants in plane the perpendicular to line of dislocation increases only linearly. Thus, in transition to 
the liquid state at point higher than the temperature of melting, a vacancy mechanism becomes so perfect, that a solid 
loses resistance to the change of form fully. And the vacancy emptiness is uniformly distributed (dissolves) in the bulk 
of material so, that a single vacancy can not be practically identified and its presence is talked about only legalistically.

For a more deep accordance it is necessary to take into account the fact of temperature dependence of the shear modulus 
in expansion of the internal energy (\ref{b30}), (\ref{b31}). If we substitute relationships (\ref{b30}) – (\ref{b32}) 
to PDF (\ref{b11}), we get a more complicated dependence of PDF on temperature in the presence of double exponents 
there. However, this does not at all affect the calculations of equilibrium values of point defects concentration in 
solids being in the free state, because in this state $\epsilon_{ij}^e=0$. Due to it, it becomes possible to examine 
the phase transition and the change of elastic properties (shear modulus) during such transition separately. 
In the presence of external stresses and the resulting elastic deformations, the influence of defect mobility must 
already be taken into account, as well as to take into account the influence of other types of defects on the process.

Earlier in a phenomenological variant, the approach was applied for the description of processes of defect formation 
and work hardening in metals treated by severe plastic deformation \cite{m10,m07,m07a,m08}, auto-vibration transitions 
between the amorphous and nanocrystalline states in amorphous alloys \cite{mm10}, processes of slipping and stick-slip 
in the super-thin lubricants \cite{mkl11,lkm11}. In particular, within the indicated models, to describe the stick-slip 
effects the processes of transitions between solid and liquid states, that is, processes of melting and solidification, 
are examined. In prospect, the method can be applied for the description of kinetic processes of defect formation and 
forming the properties in different physical systems, including soft-matter \cite{kg10,shgr11}, complex protein 
compounds \cite{ckk08}, etc.

\section{SHORT CONCLUSION}

In the article, within the framework of ideas of non-equilibrium evolutional thermodynamics, the vacancy theory 
of melting of solids is considered. With the account of removal of degeneration of vacancy energies due to their 
interaction and approximation of this interaction by polynomial presentation of the internal energy it was possible 
to generalize Frenkel concepts and to describe the process of melting as jump-like 1st-order phase transition. 
It is evident that the influence of interstitial atoms on the process of melting is not of principle, as the 
corresponding material curves lie higher than the same curves for vacancies and do not cross the fundamental 
curve in this temperature region. They can only a bit shift the proper stationary (equilibrium) solutions for 
vacancies. Influence of dislocations on the behavior of the system can play a role at relatively low temperatures. 
At temperatures close to the melting point, because of the size effect of increasing vacancy mobility, they become 
decisive in shear deformation of the material. The shear modulus here tends to zero. It is shown that apart from the 
tending to the zero of the shear modulus of the solid phase, the system has another state with the shear modulus 
identically equal to zero, which is realized exceptionally for the liquid state. The transition to this state can be 
realized earlier, than the shear modulus reaches zero, that is, the real melting point will lie somewhat to the left 
from the point at which the shear modulus becomes zero.

In addition, in the article on the example of solid with vacancies, a more general theory of non-equilibrium 
evolutional thermodynamics has been illustrated. It is shown that kinetic equations can be written not only in terms 
of the free energy, but also in terms of internal and modified free energy. The whole system of interconnected 
non-equilibrium thermodynamic potentials is obtained, including, effective internal and modified free energy. 
Intercommunication between classical and modified free energy through the component of bound energy in the both 
forms is analyzed (fig. \ref{f5}). 

\begin{acknowledgments}
Work is supported by the Budget Topic No 0109U006004 of NAS of Ukraine. 
The author thanks also Prof. V.D. Natsik for a fruitful question put at the conference, 
which stimulated additional researches of the evolution of shear modulus at melting.
\end{acknowledgments}

\end{document}